\documentclass[12pt]{article}

\usepackage{graphicx}%
\usepackage{multirow}%
\usepackage{amsmath,amssymb,amsfonts}%
\usepackage{amsthm}%
\usepackage{mathrsfs}%
\usepackage[title]{appendix}%
\usepackage{xcolor}%
\usepackage{textcomp}%
\usepackage{manyfoot}%
\usepackage{booktabs}%
\usepackage{algorithm}%
\usepackage{algorithmicx}%
\usepackage{algpseudocode}%
\usepackage{listings}%

\usepackage{hyperref}
\usepackage{times}
\usepackage{multirow}
\usepackage{geometry}
\usepackage{bm}
\usepackage{bbm}

\if@Spr@basic@refstyle%
\usepackage[authoryear]{natbib}%
\def\bibfont{\reset@font\fontfamily{\rmdefault}\normalsize\selectfont}%
\fi%
\if@Mathphys@refstyle%
\usepackage[numbers,sort&compress]{natbib}%
\def\bibfont{\reset@font\fontfamily{\rmdefault}\normalsize\selectfont}%
\fi%
\if@APS@refstyle%
\usepackage[numbers,sort&compress]{natbib}%
\def\bibfont{\reset@font\fontfamily{\rmdefault}\normalsize\selectfont}%
\fi%
\if@Vancouver@refstyle%
\usepackage[numbers,sort&compress]{natbib}%
\def\bibfont{\reset@font\fontfamily{\rmdefault}\normalsize\selectfont}%
\fi%
\if@APA@refstyle%
\usepackage[natbibapa]{apacite}%
\def\refdoi#1{\urlstyle{rm}\url{#1}}%
\AtBeginDocument{%
}%
\def\bibfont{\reset@font\fontfamily{\rmdefault}\normalsize\selectfont}%
\fi%
\if@Chicago@refstyle%
\usepackage[authoryear]{natbib}%
\hypersetup{urlcolor=black,colorlinks=false,pdfborder={0 0 0}}
\def\bibfont{\reset@font\fontfamily{\rmdefault}\normalsize\selectfont}%
\fi%
\if@Standard@Nature@refstyle%
\usepackage[numbers,sort&compress]{natbib}%
\def\bibfont{\reset@font\fontfamily{\rmdefault}\normalsize\selectfont}%
\fi%
\if@Default@refstyle%
\usepackage[numbers,sort&compress]{natbib}%
\def\bibfont{\reset@font\fontfamily{\rmdefault}\normalsize\selectfont}%
\fi%

\AtBeginDocument{\allowdisplaybreaks}%

\def\eqnheadfont{\reset@font\fontfamily{\rmdefault}\fontsize{16}{18}\bfseries\selectfont}%

\newenvironment{sciabstract}{%
\begin{quote} \bf\normalsize}
{\end{quote}}

\title{Quantum Lotka-Volterra dynamics}

\author{Yuechun Jiao$^{1,3}$, Yu Zhang$^{1}$, Jingxu Bai$^{1,3}$,Weilun Jiang$^{2,3}$, \\  
Yunhui He$^{1}$, Heng Shen$^{2,3\ast}$, Suotang Jia$^{1,3}$, Jianming Zhao$^{1,3\ast}$, C. Stuart Adams$^{4\ast}$\\
\normalsize{$^{1}$State Key Laboratory of Quantum Optics and Quantum Optics Devices, } \\
\normalsize{Institute of Laser Spectroscopy, Shanxi University, Taiyuan, 030006, China }\\ 
\normalsize{$^{2}$Institute of Opto-Electronics, Shanxi University, Taiyuan, 030006, China }\\
\normalsize{$^{3}$Collaborative Innovation Center of Extreme Optics}\\
\normalsize{Shanxi University, Taiyuan, 030006, China }\\
\normalsize{$^{4}$Joint Quantum Centre (JQC) Durham-Newcastle,}\\ 
\normalsize{Department of Physics, Durham University, DH1 3LE, United Kingdom}\\
\normalsize{$^\ast$Corresponding authors; e-mail: } \\
\normalsize{hengshen@sxu.edu.cn, zhaojm@sxu.edu.cn, c.s.adams@durham.ac.uk}
}

\date{}

\begin{document}
\maketitle
\begin{sciabstract}

Physical systems that display competitive non-linear dynamics have played a key role in the development of mathematical models of Nature.
Important examples include predator-prey models in ecology \cite{Lotka1910,VOLTERRA1926}, biology \cite{soloveichik2010},
consumer-resource models in economics \cite{Goodwin1982,Orlando2021}, and reaction-diffusion equations in chemical reactions \cite{Petrov1993,Sakata2022}.
However, as real world systems are embedded in complex environments, where it is difficult or even impossible to control external parameters, quantitative comparison between measurements and simple models remains challenging. This motivates the search for competitive dynamics in isolated physical systems, with precise control. An ideal candidate is laser excitation in dilute atomic ensembles. For example, atoms in highly-excited Rydberg states display rich many-body dynamics including ergodicity breaking \cite{Ding2024}, synchronisation \cite{Wadenpfuhl2023} and time crystals \cite{Wu2024}. Here, we demonstrate predator-prey dynamics by laser excitation and ionisation of Rydberg atoms in a room temperature vapour cell.
Ionisation of excited atoms produce electric fields that suppress further excitation. This starves the ionisation process of resource, giving rise to predator-prey dynamics. By comparing our results to the Lotka-Volterra model, we demonstrate that as well applications in non-linear dynamics, our experiment has applications in metrology, and remote sensing of localised plasmas.

\end{sciabstract}

The Lotka-Volterra equations~\cite{Lotka1910,VOLTERRA1926} provide a paradigmatic model of coupled competitive non-linear dynamics. Although most frequently used to describe predator-prey dynamics in ecological communities~\cite{goel1971volterra,altieri2021properties,sidhom2020ecological}, generalised Lotka-Volterra equations have also been applied to chemical reactions \cite{Lotka1910}, phase transitions, fluctuations, and multi-stability in driven-dissipative bosonic systems ~\cite{Satulovsky1994,Mobilia2006,Antal2001,knebel2015evolutionary}, topological phases ~\cite{knebel2020topological,yoshida2021chiral} and non-Hermitian physics~\cite{zhang2023emergent}. However, direct and robust observations of Lotka-Volterra dynamics remain elusive. Here, we report a realization of predator-prey like dynamics in a dilute atomic gas. A direct quantum interface to light emerges as a powerful and ubiquitous tool, enabling the precise control of all external parameters in unprecedented ways compared to classical systems.  

\begin{figure*}
\centering
\includegraphics[width=\linewidth]{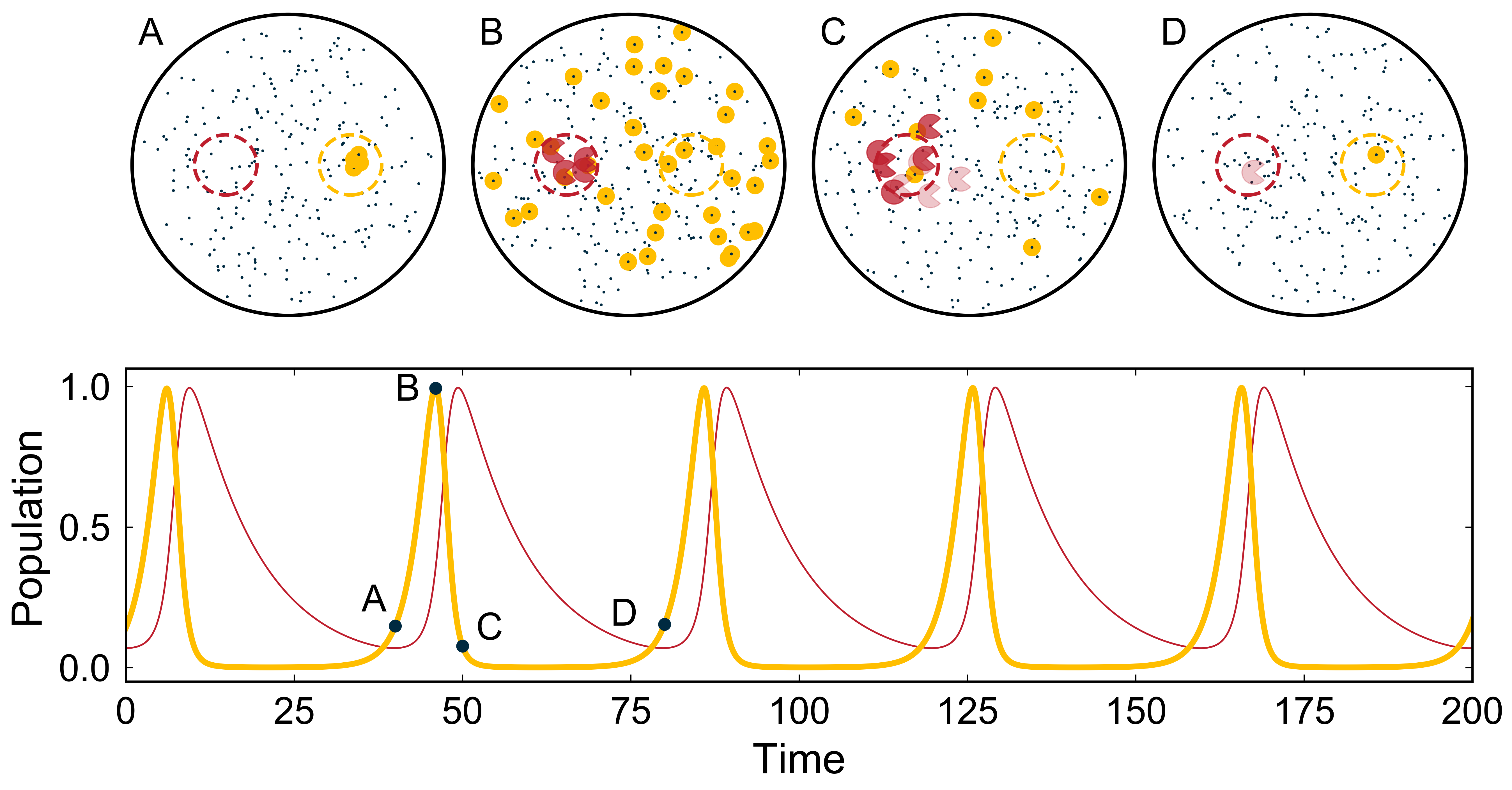}
\caption{\textbf{Quantum Lotka-Volterra dynamics.}
{\bf A--D}: End view of the cylindrical vapour cell. Atoms, shown as black dots are everywhere. Rydberg states and charges are shown in yellow and red, respectively. The excitation and ionising lasers are applied in the regions indicated by the yellow and red dotted circles, respectively. The size of the laser beams relative to the size of the cell is greatly exaggerated. 
{\bf A} Beginning of the cycle. There are few or no charges in the cell. The excitation lasers produce growth in the number of ground-Rydberg superpositions, yellow circles with a black dot. {\bf B} The ground-Rydberg superposition atoms diffuse throughout the cell. Some are photo-ionised in the region of the ionising laser or ionised via collision with existing charges. The resulting charges are shown in red. {\bf C} The charges produce an electric field which shifts the Rydberg energy level preventing further excitation of the Rydberg state. Deprived of Rydbergs, the number of ions begins to decay. {\bf D} Once the number of ions falls below a critical threshold the electric fields are sufficiently diminished that Rydberg excitation is permitted again, and the cycle repeats. Below we show the solution of equations (\ref{eq:lv1}) and (\ref{eq:lv2}), the prey, $x$ shown in yellow, and the predator, $y$, shown in red. Approximate positions in the cycles corresponding to A, B, C and D are indicated. }
\label{fig:idea}
\end{figure*}

We begin by introducing the two-species Lotka-Volterra equations,
\begin{eqnarray}
    \frac{{\rm d}x}{{\rm d}t} & = & \alpha~ x - \beta~x y~,\label{eq:lv1}\\
    \frac{{\rm d}y}{{\rm d}t} & = & -\gamma~ y + \delta~ x y~,\label{eq:lv2} 
\end{eqnarray}
where $x$ and $y$ are populations, and
$\alpha$, $\beta$, $\gamma$ and $\delta$ are positive constants. Initially, species $x$---the prey---feeds on an external resource and can grow exponentially at a rate given by $\alpha$. Species $y$---the predator---feeds on $x$. Consequently, $x$ is depleted at rate proportional to $y$ with constant of proportionality $\beta$. The predator population $y$ grows at the rate proportional to $x$ with constant of proportionality $\delta$, and decays at a rate proportional to $\gamma$. In our experiment, the prey and predator are Rydberg atoms---or rather, atoms in a superposition of ground and Rydberg states---and charges, ions and electrons, respectively.
The experimental scenario is depicted schematically in Fig.~\ref{fig:idea}. A cylindrical glass cell contains a vapour of caesium (Cs) atoms. A probe and coupling laser excites atoms into a superposition of ground and Rydberg states in the region bounded by the yellow dotted circle. The Rydberg states are relatively long lived (up to 100 $\mu$s). Consequently, the ground-Rydberg superpositions---shown as black dots in a yellow circle in Fig.~\ref{fig:idea}---can diffuse throughout the cell. A second photoionising laser, in the region indicated by the red dotted circle in Fig.~\ref{fig:idea}, ionises the Rydberg component. The resulting charged particles are shown in red in Fig.~\ref{fig:idea}. The number of charges can also increase via collisions with atoms in Rydberg superposition states. As the ions and electrons produce significant electric fields, they change the resonance condition for the excitation which cuts off the supply of Rydberg excitations. This electromagnetic feedback mechanism is somewhat analogous to synchronisation of fireflies which also interact via light \cite{Buck1988}. 
Subsequently, deprived of `feed', the population of ions and electrons is depleted via recombination and collisions with the walls of the cell. Once the number of charges falls below a critical threshold, the cycle can begin again. We can expect both the Rydberg and charge populations to oscillate as shown in lower plot in Fig.~\ref{fig:idea}. Although, a complete model of the dynamics would require knowledge of ion and electron motion, induced electric fields, atom-charge collisions, etc, we find excellent agreement with the relatively simple Lotka-Volterra equations, (\ref{eq:lv1}) and (\ref{eq:lv2}). However, before presenting the experimental data, we need to explain the experiment and key parameters in more detail.

\begin{figure}
\centering
\includegraphics[width=\linewidth]{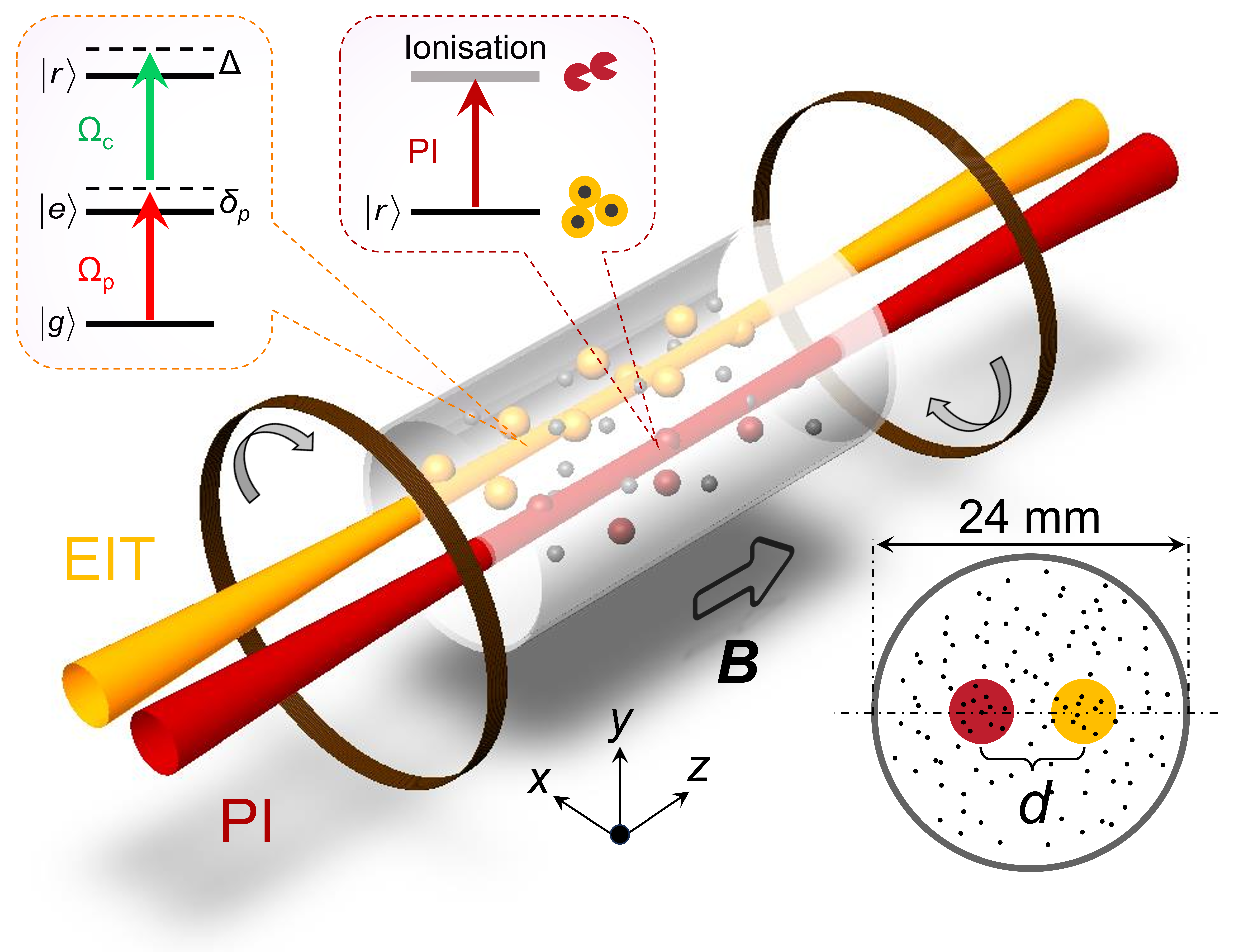}
\caption{\textbf{Experiment.} Rydberg lasers (yellow) consisting of a probe and a coupling beam counterpropagating through a Cs vapor cell to form EIT. A PI laser (red) co-propagates parallel with the coupling laser, ionising the Rydberg atoms, see the level scheme in the top panel. A homogeneous magnetic field \textit{B} is applied along with the probe direction, generated with a pair of Helmholtz coils. The probe and coupling lasers excite the transition of $|g\rangle$ $\to$ $|r\rangle$ via the intermediate state $|e\rangle$ with detuning of $\delta_p = 2\pi \times 110$~MHz. Black, yellow and red dots represent the ground state atoms, Rydberg states and charges, respectively. Inset: End view of the cylindrical vapour cell, where Rydberg excitation lasers (yellow solid circle) and PI laser (red solid circle) is separated by a distance of $d$.}
\label{fig:experiment}
\end{figure}

We consider an ensemble of atoms in a room temperature vapor cell. As illustrated in Fig.~\ref{fig:experiment}, Rydberg excitation lasers (marked yellow laser in the cell), referred as a probe (852 nm, $\Omega_p$) and a coupling laser (509 nm, $\Omega_c$) with $\sigma^+$ circular polarization, are overlapped and counterpropagate through the cell. By measuring the probe transmission we observe a Rydberg electromagnetically induced transparency (EIT) resonance \cite{Mohapatra2017}.
The probe laser is locked to the transition of $|g\rangle$ $\to$ $|e\rangle$ with detuning of $\delta_p$, while the coupling laser drives the transition of $|e\rangle$ $\to$ $|r\rangle$. A 
photo-ionisation (PI) laser beam also with wavelength 509~nm is co-propagated in parallel through the cell at a distance of $d$ relative to the coupling laser. The PI laser, marked red laser in the cell in Fig.~\ref{fig:experiment},  has a power up to 200~mW which is sufficient to ionise atoms in the Rydberg state that cross its path. Independent tests of ionisations and the time delays associated with atomic motion are described in the Methods section. A homogeneous magnetic field, \textit{B}, is applied parallel to the probe direction. A finite magnetic field is found to be essential to reduce the charge decay rate, the coefficient $\gamma$ in equation (\ref{eq:lv2}), to a level where oscillation are possible, see Methods. The inset shows the end view of the cylindrical vapour cell indicating the distance $d$ between Rydberg excitation lasers (yellow solid circle) and PI laser (red solid circle).

\begin{figure}
\centering
\includegraphics[width=\linewidth]{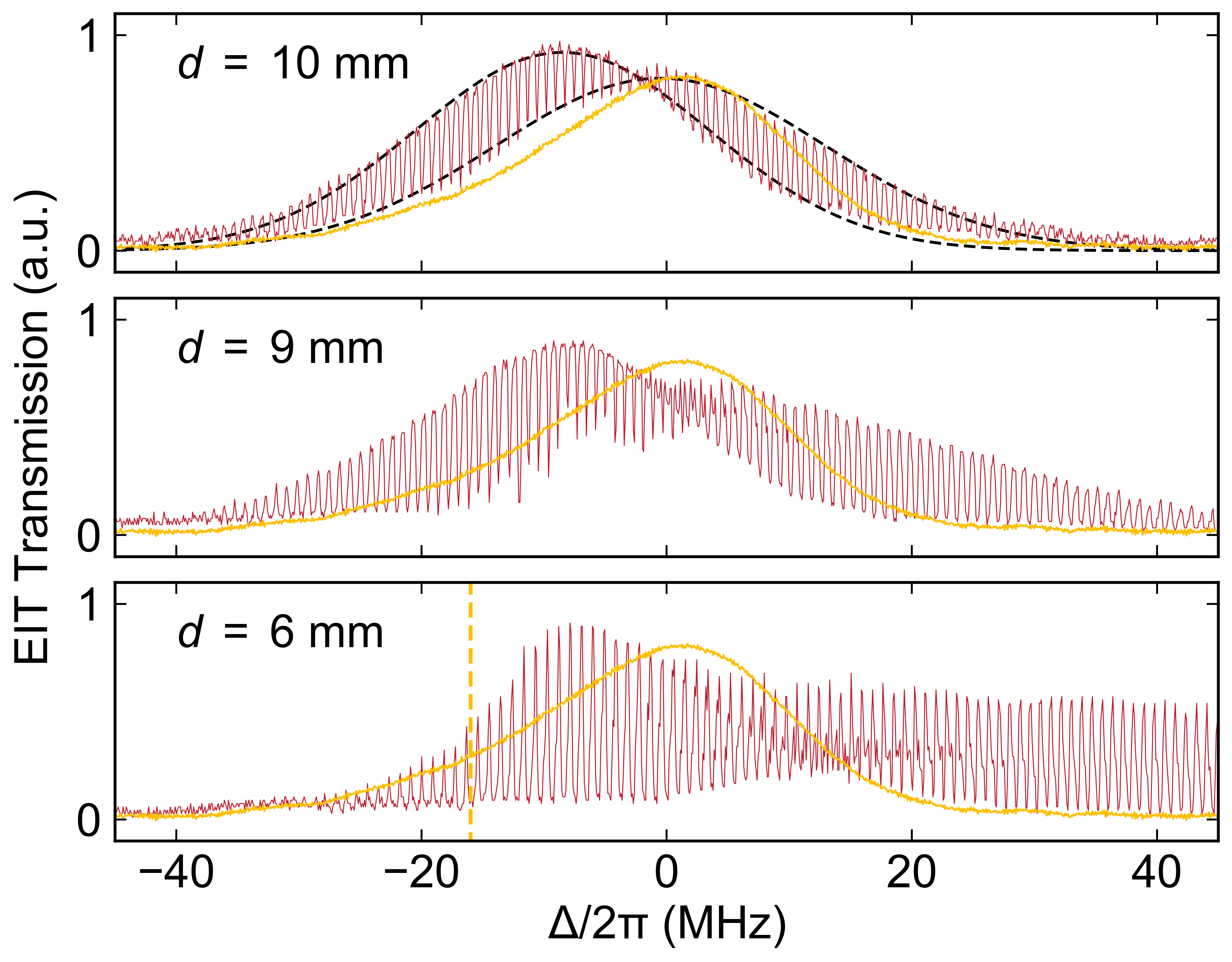}
\caption{\textbf{EIT Spectra.} The Rydberg EIT 
spectra without (yellow) and with (red) the additional PI beam. For this example with magnetic field $B=11.6$~G, the unperturbed EIT linewidth is 12~MHz.
The beam separations are $d = 10$~mm, 9 mm and 6 mm. For $d = 10$~mm, we see that the transmission oscillates between two lineshapes indicated by the dashed black lines. These two lineshapes corresponding to the case of low and high charge in the cell. The shift in the resonance due to the charges is 9~MHz. For smaller beam spacing, $d$, the spectrum is more complex, however, we can still characterise the effect as switching between state with low and high charge. The yellow dashed line indicates the lock frequency used in the later measurements.}
\label{fig:spectra}
\end{figure}

To experimentally observe the predator-prey like dynamic, we measure the influence of the PI laser on the Rydberg EIT spectra. These measurements are performed with Rabi frequency of probe and coupling lasers given by $\Omega_p = 2 \pi \times 25.10$~MHz and $\Omega_c = 2 \pi \times0.59$~MHz, respectively, and an axial magnetic field, \textit{B} = 11.6~G. In Fig.~\ref{fig:spectra}, we plot the EIT spectra without and with the PI laser at beam separations of $d = 6$~mm, 9~mm and 10~mm. When the PI laser is switched off we observe a normal EIT resonance spectrum with a gaussian lineshape with width $12$~MHz, yellow lines in Fig.~\ref{fig:spectra}. When the PI laser is turned on, we observe periodic oscillations superimposed onto the EIT spectrum. Apparent in the $d = 10$~mm spectrum is that the probe transmission is rapidly switching between two curves---the unperturbed resonance where the number of charges is negligible and a shifted resonance where the effect of ionisation is significant.

\begin{figure*}
\centering
\includegraphics[width=\linewidth]{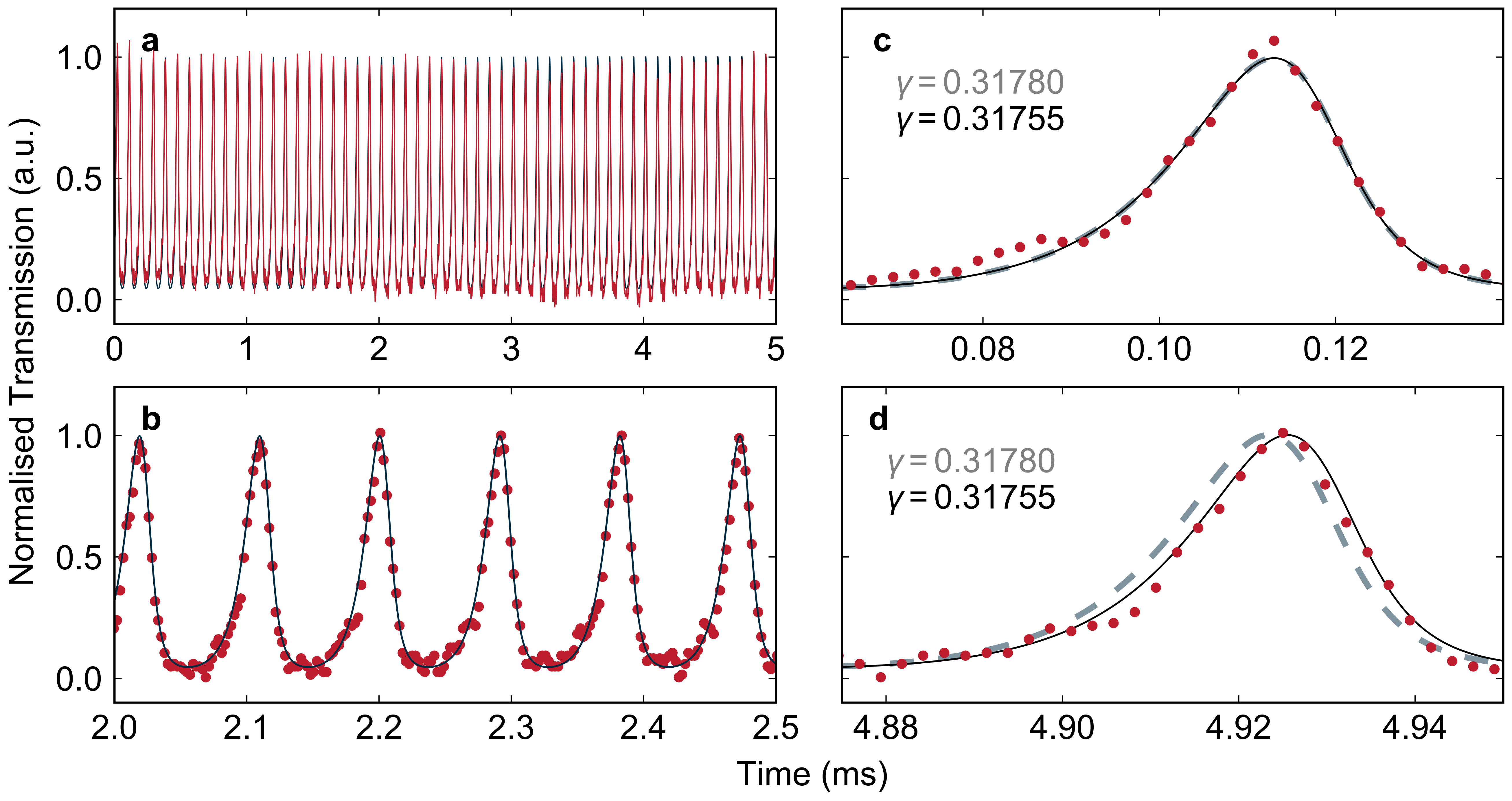}
\caption{\textbf{Lotka-Volterra fit.} \textbf{a, b,} Black and red show the prediction of the Lotka-Volterra model, equations (\ref{eq:lv1}) and (\ref{eq:lv2}), and the experimental data, respectively.
Probe transmission measured over 5~ms for \textbf{a}, and a 0.5~ms time slice for \textbf{b}. We see how both the rising and fall edge of the data are accurately reproduced by the Lotka-Volterra equations. 
The parameters of the fit are $\alpha=0.75$, $\beta=0.25$, $\gamma=0.31755$, and $\delta= 0.25$ in units of milliseconds. To obtain an accurate fit over 5 ms requires that the parameter $\gamma$ in equation (\ref{eq:lv2}) is specified to better than a part in one thousand. This is illustrated in
\textbf{c, d,} the first and last pulse of the 5~ms time sequence shown in \textbf{a}. The Lotka-Volterra curves for $\gamma=0.31755$ and $0.31780$ are shown using a solid black line and a grey dashed line, respectively. The other parameters in this plot are: distance between the probe and ionisation region is $d = 6$~mm, excitation lasers locked with a detuning, $\Delta = -2\pi \times 8$~MHz, probe and coupling laser powers 200~$\mu$W and 3 mW, respectively, photo-ionisation laser power 200 mW, and magnetic field
$B=11.6$~G.
}
\label{fig:LV_data}
\end{figure*}

To investigate these EIT oscillations, we lock the coupling laser to a detuning $\Delta = - 2\pi \times 8$~MHz (where $\Delta$ is defined by the detuning relative to the peak of each spectrum, dashed yellow line in Fig.~\ref{fig:spectra}), and observe the probe transmission as a function of time, see Extended Data Fig.~2.  In Fig.~\ref{fig:LV_data} we compare a typical experimental time sequence with the prediction of the Lotka-Volterra model, equations (\ref{eq:lv1}) and (\ref{eq:lv2}). This comparison illustrates how the model accurately reproduces both the long and short time dynamics, see  Fig.~\ref{fig:LV_data}a and b, respectively, and also Extended Data Fig.~3. By measuring over many cycles we can obtain a very accurate estimate of the parameters $\alpha$, $\beta$, $\gamma$ and $\delta$. In particular, in order to fit the data over a 5~ms sequence, Fig.~\ref{fig:LV_data}a, we need to specify the Lotka-Volterra parameter $\gamma$ to better than one part in a thousand. To illustrate this sensitivity, in Fig.~\ref{fig:LV_data}c and d we show the first and last pulse in a 5~ms time sequence. For the last pulse, the difference between a plasma decay constant $\gamma=0.31755$ and $\gamma=0.31780$ (a difference of 8 part in $10^4$) is significant. This suggests that the observed oscillations could be useful in both metrology and quantum sensing applications. By turning a constant observable into a frequency measurement, higher precision is achieved. This adds a potentially attractive alternative to the already powerful capabilities of Rydberg atom sensors \cite{Adams2020,Ding2022,Borowka2024} 

In summary, we have demonstrated predator-prey dynamics in an atomic system using laser excitation and ionisation. The dynamics are probed using EIT, which
detects the quantum coherence of atoms in a superposition of ground and excited states. We show that the time evolution of the EIT signal is accurately described by the Lotka-Volterra equations. As we have precise control over experimental parameters our technique is able to provide robust complex-system data that could be used to train deep learning models \cite{Bury2023}.
In addition, as the Lotka-Volterra cycle is extremely sensitive to external fields our device could be used for precision measurement of static or low frequency electromagnetic fields.

\bibliographystyle{sn-nature}
\bibliography{references}

\begin{thebibliography}{10}
\expandafter\ifx\csname url\endcsname\relax
  \def\url#1{\burl{#1}}\fi
\expandafter\ifx\csname urlprefix\endcsname\relax\def\urlprefix{URL }\fi
\providecommand{\bibinfo}[2]{#2}
\providecommand{\eprint}[2][]{\url{#2}}
\providecommand{\doi}[1]{\url{https://doi.org/#1}}
\bibcommenthead

\bibitem{Lotka1910}
\bibinfo{author}{Lotka, A.~J.}
\newblock \bibinfo{title}{Contribution to the theory of periodic reactions}.
\newblock \emph{\bibinfo{journal}{J. Phys. Chem.}} \textbf{\bibinfo{volume}{14}}, \bibinfo{pages}{271--274} (\bibinfo{year}{1910}).

\bibitem{VOLTERRA1926}
\bibinfo{author}{Volterra, V.}
\newblock \bibinfo{title}{Fluctuations in the abundance of a species considered mathematically1}.
\newblock \emph{\bibinfo{journal}{Nature}} \textbf{\bibinfo{volume}{118}}, \bibinfo{pages}{558--560} (\bibinfo{year}{1926}).

\bibitem{soloveichik2010}
\bibinfo{author}{Soloveichik, D.}, \bibinfo{author}{Seelig, G.} \& \bibinfo{author}{Winfree, E.}
\newblock \bibinfo{title}{{DNA} as a universal substrate for chemical kinetics}.
\newblock \emph{\bibinfo{journal}{Proc. Natl. Acad. Sci. U.S.A.}} \textbf{\bibinfo{volume}{107}}, \bibinfo{pages}{5393--5398} (\bibinfo{year}{2010}).

\bibitem{Goodwin1982}
\bibinfo{author}{Goodwin, R.~M.}
\newblock \emph{\bibinfo{title}{A Growth Cycle}}, \bibinfo{pages}{165--170} (\bibinfo{publisher}{Palgrave Macmillan UK}, \bibinfo{address}{London}, \bibinfo{year}{1982}).

\bibitem{Orlando2021}
\bibinfo{author}{Orlando, G.} \& \bibinfo{author}{Sportelli, M.}
\newblock \emph{\bibinfo{title}{Growth and Cycles as a Struggle: Lotka--Volterra, Goodwin and Phillips}}, \bibinfo{pages}{191--208} (\bibinfo{publisher}{Springer International Publishing}, \bibinfo{address}{Cham}, \bibinfo{year}{2021}).

\bibitem{Petrov1993}
\bibinfo{author}{Petrov, V.}, \bibinfo{author}{G{\'a}sp{\'a}r, V.}, \bibinfo{author}{Masere, J.} \& \bibinfo{author}{Showalter, K.}
\newblock \bibinfo{title}{Controlling chaos in the belousov---zhabotinsky reaction}.
\newblock \emph{\bibinfo{journal}{Nature}} \textbf{\bibinfo{volume}{361}}, \bibinfo{pages}{240--243} (\bibinfo{year}{1993}).

\bibitem{Sakata2022}
\bibinfo{author}{Sakata, T.} \emph{et~al.}
\newblock \bibinfo{title}{Self-oscillating chemoelectrical interface of solution-gated ion-sensitive field-effect transistor based on belousov--zhabotinsky reaction}.
\newblock \emph{\bibinfo{journal}{Sci. Rep.}} \textbf{\bibinfo{volume}{12}}, \bibinfo{pages}{2949} (\bibinfo{year}{2022}).

\bibitem{Ding2024}
\bibinfo{author}{Ding, D.} \emph{et~al.}
\newblock \bibinfo{title}{Ergodicity breaking from rydberg clusters in a driven-dissipative many-body system}.
\newblock \emph{\bibinfo{journal}{Sci. Adv.}} \textbf{\bibinfo{volume}{10}}, \bibinfo{pages}{eadl5893} (\bibinfo{year}{2024}).

\bibitem{Wadenpfuhl2023}
\bibinfo{author}{Wadenpfuhl, K.} \& \bibinfo{author}{Adams, C.~S.}
\newblock \bibinfo{title}{Emergence of synchronization in a driven-dissipative hot rydberg vapor}.
\newblock \emph{\bibinfo{journal}{Phys. Rev. Lett.}} \textbf{\bibinfo{volume}{131}}, \bibinfo{pages}{143002} (\bibinfo{year}{2023}).

\bibitem{Wu2024}
\bibinfo{author}{Wu, X.} \emph{et~al.}
\newblock \bibinfo{title}{Dissipative time crystal in a strongly interacting rydberg gas}.
\newblock \emph{\bibinfo{journal}{Nat. Phys.}}  (\bibinfo{year}{2024}).

\bibitem{goel1971volterra}
\bibinfo{author}{Goel, N.~S.}
\newblock \emph{\bibinfo{title}{On the {{Volterra}} and Other Nonlinear Models of Interacting Populations [by] {{N}}.{{S}}. {{Geol}}, {{S}}.{{C}}. {{Maitra}} [and] {{E}}.{{W}}. {{Montroll}}.}}  (\bibinfo{publisher}{Academic Press}, \bibinfo{address}{New York}, \bibinfo{year}{1971}).

\bibitem{altieri2021properties}
\bibinfo{author}{Altieri, A.}, \bibinfo{author}{Roy, F.}, \bibinfo{author}{Cammarota, C.} \& \bibinfo{author}{Biroli, G.}
\newblock \bibinfo{title}{Properties of {{Equilibria}} and {{Glassy Phases}} of the {{Random Lotka-Volterra Model}} with {{Demographic Noise}}}.
\newblock \emph{\bibinfo{journal}{Phys. Rev. Lett.}} \textbf{\bibinfo{volume}{126}}, \bibinfo{pages}{258301} (\bibinfo{year}{2021}).

\bibitem{sidhom2020ecological}
\bibinfo{author}{Sidhom, L.} \& \bibinfo{author}{Galla, T.}
\newblock \bibinfo{title}{Ecological communities from random generalized {{Lotka-Volterra}} dynamics with nonlinear feedback}.
\newblock \emph{\bibinfo{journal}{Phys. Rev. E}} \textbf{\bibinfo{volume}{101}}, \bibinfo{pages}{032101} (\bibinfo{year}{2020}).

\bibitem{Satulovsky1994}
\bibinfo{author}{Satulovsky, J.~E.} \& \bibinfo{author}{Tom\'e, T.}
\newblock \bibinfo{title}{Stochastic lattice gas model for a predator-prey system}.
\newblock \emph{\bibinfo{journal}{Phys. Rev. E}} \textbf{\bibinfo{volume}{49}}, \bibinfo{pages}{5073--5079} (\bibinfo{year}{1994}).

\bibitem{Mobilia2006}
\bibinfo{author}{Mobilia, M.}, \bibinfo{author}{Georgiev, I.~T.} \& \bibinfo{author}{T\"auber, U.~C.}
\newblock \bibinfo{title}{Fluctuations and correlations in lattice models for predator-prey interaction}.
\newblock \emph{\bibinfo{journal}{Phys. Rev. E}} \textbf{\bibinfo{volume}{73}}, \bibinfo{pages}{040903} (\bibinfo{year}{2006}).

\bibitem{Antal2001}
\bibinfo{author}{Antal, T.} \& \bibinfo{author}{Droz, M.}
\newblock \bibinfo{title}{Phase transitions and oscillations in a lattice prey-predator model}.
\newblock \emph{\bibinfo{journal}{Phys. Rev. E}} \textbf{\bibinfo{volume}{63}}, \bibinfo{pages}{056119} (\bibinfo{year}{2001}).

\bibitem{knebel2015evolutionary}
\bibinfo{author}{Knebel, J.}, \bibinfo{author}{Weber, M.~F.}, \bibinfo{author}{Kr{\"u}ger, T.} \& \bibinfo{author}{Frey, E.}
\newblock \bibinfo{title}{Evolutionary games of condensates in coupled birth--death processes}.
\newblock \emph{\bibinfo{journal}{Nat. Commun.}} \textbf{\bibinfo{volume}{6}}, \bibinfo{pages}{6977} (\bibinfo{year}{2015}).

\bibitem{knebel2020topological}
\bibinfo{author}{Knebel, J.}, \bibinfo{author}{Geiger, P.~M.} \& \bibinfo{author}{Frey, E.}
\newblock \bibinfo{title}{Topological {{Phase Transition}} in {{Coupled Rock-Paper-Scissors Cycles}}}.
\newblock \emph{\bibinfo{journal}{Phys. Rev. Lett.}} \textbf{\bibinfo{volume}{125}}, \bibinfo{pages}{258301} (\bibinfo{year}{2020}).

\bibitem{yoshida2021chiral}
\bibinfo{author}{Yoshida, T.}, \bibinfo{author}{Mizoguchi, T.} \& \bibinfo{author}{Hatsugai, Y.}
\newblock \bibinfo{title}{Chiral edge modes in evolutionary game theory: {{A}} kagome network of rock-paper-scissors cycles}.
\newblock \emph{\bibinfo{journal}{Phys. Rev. E}} \textbf{\bibinfo{volume}{104}}, \bibinfo{pages}{025003} (\bibinfo{year}{2021}).

\bibitem{zhang2023emergent}
\bibinfo{author}{Zhang, T.} \& \bibinfo{author}{Cai, Z.}
\newblock \bibinfo{title}{Emergent non-{{Hermitian}} physics in a generalized {{Lotka-Volterra}} model}.
\newblock \emph{\bibinfo{journal}{Phys. Rev. B}} \textbf{\bibinfo{volume}{108}}, \bibinfo{pages}{104304} (\bibinfo{year}{2023}).

\bibitem{Buck1988}
\bibinfo{author}{Buck, J.}
\newblock \bibinfo{title}{Synchronous rhythmic flashing of fireflies. {II}.}
\newblock \emph{\bibinfo{journal}{Q. Rev. Biol.}} \textbf{\bibinfo{volume}{63}}, \bibinfo{pages}{265--289} (\bibinfo{year}{1988}).

\bibitem{Mohapatra2017}
\bibinfo{author}{Mohapatra, A.~K.}, \bibinfo{author}{Jackson, T.~R.} \& \bibinfo{author}{Adams, C.~S.}
\newblock \bibinfo{title}{Coherent optical detection of highly excited rydberg states using electromagnetically induced transparency}.
\newblock \emph{\bibinfo{journal}{Phys. Rev. Lett.}} \textbf{\bibinfo{volume}{98}}, \bibinfo{pages}{113003} (\bibinfo{year}{2007}).

\bibitem{Adams2020}
\bibinfo{author}{Adams, C.~S.}, \bibinfo{author}{Pritchard, J.~D.} \& \bibinfo{author}{Shaffer, J.~P.}
\newblock \bibinfo{title}{Rydberg atom quantum technologies}.
\newblock \emph{\bibinfo{journal}{J. Phys. B: At. Mol. Opt. Phys.}} \textbf{\bibinfo{volume}{53}}, \bibinfo{pages}{012002} (\bibinfo{year}{2019}).

\bibitem{Ding2022}
\bibinfo{author}{Ding, D.-S.} \emph{et~al.}
\newblock \bibinfo{title}{Enhanced metrology at the critical point of a many-body rydberg atomic system}.
\newblock \emph{\bibinfo{journal}{Nat. Phys.}} \textbf{\bibinfo{volume}{18}}, \bibinfo{pages}{1447--1452} (\bibinfo{year}{2022}).

\bibitem{Borowka2024}
\bibinfo{author}{Bor{\'o}wka, S.}, \bibinfo{author}{Pylypenko, U.}, \bibinfo{author}{Mazelanik, M.} \& \bibinfo{author}{Parniak, M.}
\newblock \bibinfo{title}{Continuous wideband microwave-to-optical converter based on room-temperature rydberg atoms}.
\newblock \emph{\bibinfo{journal}{Nat. Photonics}} \textbf{\bibinfo{volume}{18}}, \bibinfo{pages}{32--38} (\bibinfo{year}{2024}).

\bibitem{Bury2023}
\bibinfo{author}{Bury, T.~M.} \emph{et~al.}
\newblock \bibinfo{title}{Predicting discrete-time bifurcations with deep learning}.
\newblock \emph{\bibinfo{journal}{Nat. Commun.}} \textbf{\bibinfo{volume}{14}}, \bibinfo{pages}{6331} (\bibinfo{year}{2023}).

\bibitem{Weller2016}
\bibinfo{author}{Weller, D.}, \bibinfo{author}{Urvoy, A.}, \bibinfo{author}{Rico, A.}, \bibinfo{author}{L\"ow, R.} \& \bibinfo{author}{K\"ubler, H.}
\newblock \bibinfo{title}{Charge-induced optical bistability in thermal rydberg vapor}.
\newblock \emph{\bibinfo{journal}{Phys. Rev. A}} \textbf{\bibinfo{volume}{94}}, \bibinfo{pages}{063820} (\bibinfo{year}{2016}).

\bibitem{Weller2019}
\bibinfo{author}{Weller, D.}, \bibinfo{author}{Shaffer, J.~P.}, \bibinfo{author}{Pfau, T.}, \bibinfo{author}{L\"ow, R.} \& \bibinfo{author}{K\"ubler, H.}
\newblock \bibinfo{title}{Interplay between thermal rydberg gases and plasmas}.
\newblock \emph{\bibinfo{journal}{Phys. Rev. A}} \textbf{\bibinfo{volume}{99}}, \bibinfo{pages}{043418} (\bibinfo{year}{2019}).

\bibitem{Killian2007}
\bibinfo{author}{Killian, T.}, \bibinfo{author}{Pattard, T.}, \bibinfo{author}{Pohl, T.} \& \bibinfo{author}{Rost, J.}
\newblock \bibinfo{title}{Ultracold neutral plasmas}.
\newblock \emph{\bibinfo{journal}{Phys. Rep.}} \textbf{\bibinfo{volume}{449}}, \bibinfo{pages}{77--130} (\bibinfo{year}{2007}).

\bibitem{May1972}
\bibinfo{author}{May, R.~M.}
\newblock \bibinfo{title}{Limit cycles in predator-prey communities}.
\newblock \emph{\bibinfo{journal}{Science}} \textbf{\bibinfo{volume}{177}}, \bibinfo{pages}{900--902} (\bibinfo{year}{1972}).

\bibitem{May1973}
\bibinfo{author}{May, R.~M.}
\newblock \bibinfo{title}{Time-delay versus stability in population models with two and three trophic levels}.
\newblock \emph{\bibinfo{journal}{Ecology}} \textbf{\bibinfo{volume}{54}}, \bibinfo{pages}{315--325} (\bibinfo{year}{1973}).

\end{thebibliography}

\section*{Methods}

\newcounter{Sfigure}
\setcounter{Sfigure}{1}
\renewcommand{\thefigure}{\arabic{Sfigure}}
\renewcommand{\figurename}{Extended Data Fig.}

\subsection*{Experimental details}

The experiment is performed in a room-temperature Cs vapour cell with a size of $\phi$ 2.4 cm $\times$ 7.5 cm. The 852~nm and 509~nm lasers are provided by Toptica DL pro and Precilasers, respectively. The probe laser is locked to the transition of $|g\rangle = |6S_{1/2}, F = 4, m_F = 4\rangle$ $\to$ $|e\rangle$ = $|6P_{3/2}, F^\prime = 5, m_F^\prime = 5\rangle$ with detuning of $\delta_p$ = 2$\pi \times$110~MHz, while the coupling laser is either scanned through a two-photon resonance with the $|r\rangle$ = $|60D_{5/2}, m_j = 5/2\rangle$ Rydberg state, or locked at a fixed laser detuning relative to resonance. A high finesse ultralow expansion (ULE) cavity provides the feedback for the frequency stabilization of both probe and coupling lasers (FSR: 1.5 GHz, Finesse $1.5\times 10^4$). The tunable frequency offset locking technique ensures the measurement of quantum Lotka-Volterra dynamics with arbitrary laser detuning. The $1/e^2$ beam waist of 852~nm and 509~nm are $w_p$ = 500~$\mu$m and $w_c$ = 425~$\mu$m, respectively. 
The PI laser beam also with wavelength 509~nm but shifted $-110$~MHz by an AOM relative to the coupling laser is co-propagated in parallel through the cell. The homogeneous magnetic field \textit{B}, generated by a pair of Helmholtz coils with 220~mm inner diameter, is applied parallel to the probe direction.

\subsection*{Transit of Rydberg atoms and ionisation}

\begin{figure*}[htbp]
    \centering
    \includegraphics[width=0.9\textwidth]{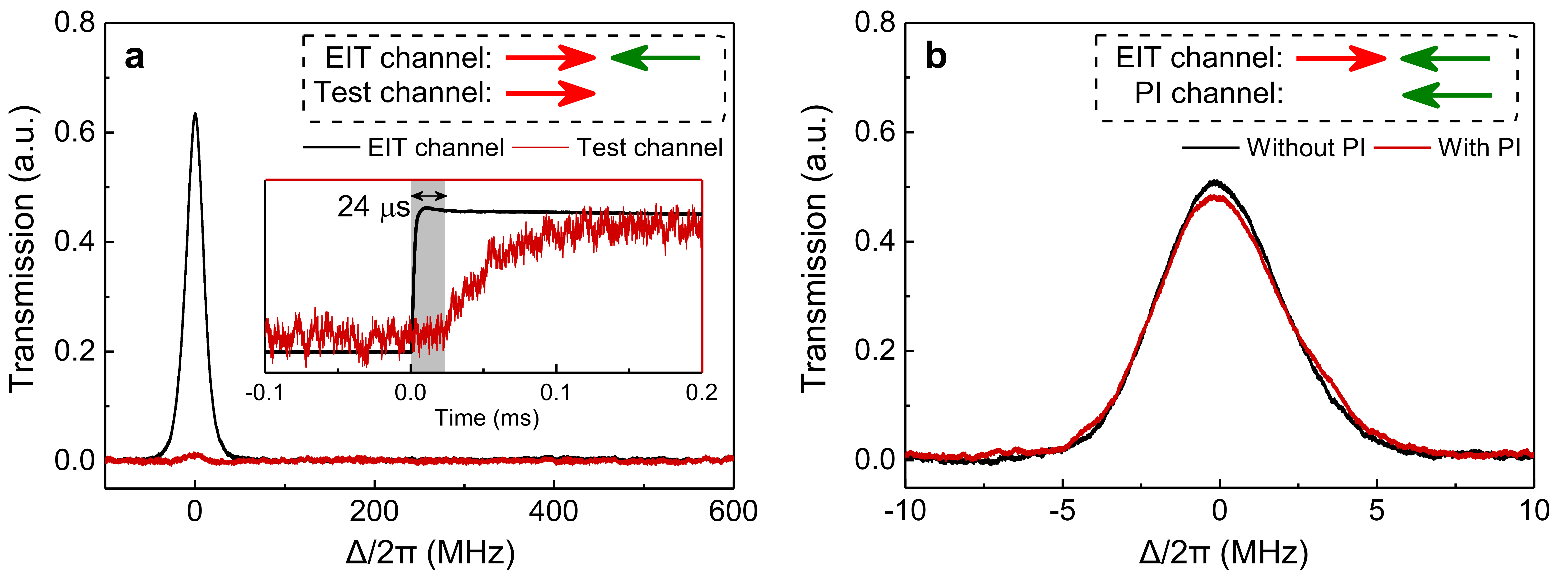}
    \caption{\textbf{Transit of Rydberg atoms and ionisation.} \textbf{a,} Measurement of transmission of the EIT probe laser and test laser. Inset show the establishment of EIT and transmission of the test laser with pulsed EIT coupling laser. \textbf{b,} Demonstration of change of EIT spectra in the presence of the PI laser.}
    \label{fig:transit}
    \addtocounter{Sfigure}{1}
\end{figure*}

After the probe and coupling lasers excite ground atoms to Rydberg states, the Rydberg atoms diffuse throughout the cell due to their long lifetime. We then apply a test 852~nm laser (same frequency as the EIT probe laser), which measures the local transmission through the cell. The test laser co-propagates with the EIT probe with a transverse displace $d$ equal to 6~mm. If the atoms in a ground-Rydberg superposition transit the test laser region, they do not scatter resulting in an enhanced transmission of the test laser. In Extended Data Fig.~\ref{fig:transit}a, we demonstrate the signal in EIT channel and test channel, which show the test laser has an enhanced transmission. We use this technique to measure the time scale of atom transport across the cell. The EIT coupling laser is pulsed with a 50~Hz square modulation. Both the probe and coupling lasers are locked. The inset shows the onset of EIT and the corresponding response of the test laser. The results show that the enhanced transmission of the test laser occurs after a 24~$\mu$s delay. This is consistent with the transit time that Rydberg atoms travel to the test laser region.

We then measure the influence of the PI laser on the EIT without the magnetic field \textit{B}. In Extended Data Fig.~\ref{fig:transit}b, we demonstrate EIT spectra without and with PI laser. The results show the EIT peak is reduced in the presence of the PI laser. This verifies the generation of the ions.

\subsection*{Time dynamics of the probe transmission}

\begin{figure*}
\centering
\includegraphics[width=\linewidth]{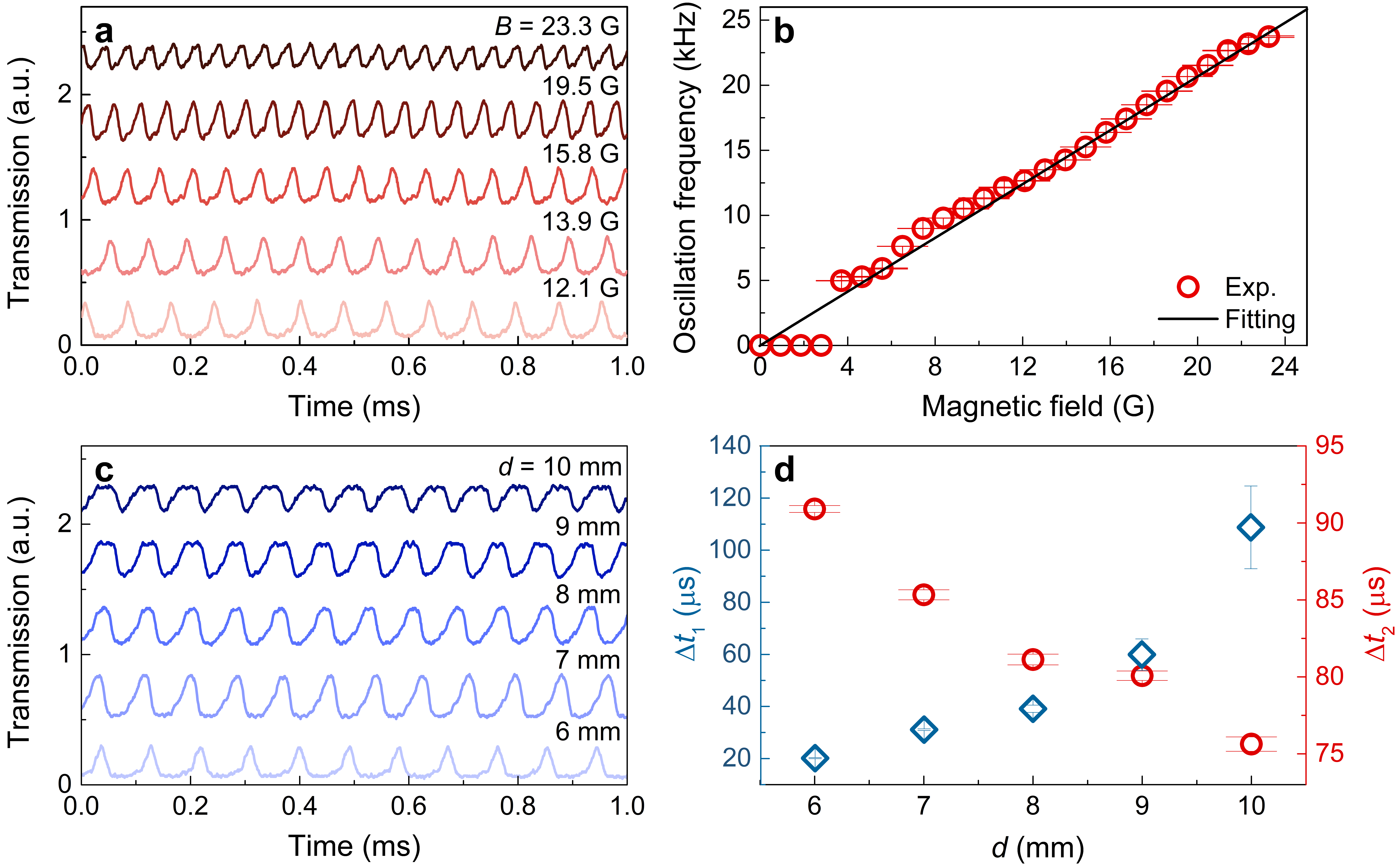}
\caption{\textbf{Time dynamics of the probe transmission.}
\textbf{a,} The measured Lotka-Volterra cycles at different value of the external magnetic field \textit{B}. \textbf{b,} The frequency of the oscillations as a function of the magnetic field, showing the threshold for plasma formation. A solid line displays the linear fitting to the data, corresponding slope is 1.033 $\pm$ 0.006~kHz/G. The error bars show the standard deviation of three independent measurements. \textbf{c,} The measured Lotka-Volterra cycle at different distances, $d$. \textbf{d,} The linewidth of the pulses ($\Delta t_{1}$) and the time between pulses ($\Delta t_{2}$) as a function of the distance, $d$. The error bars show the standard deviation of three independent measurements.}
\label{fig:plasma} \addtocounter{Sfigure}{1}
\end{figure*}

Typical experimental time sequences as a function of both the magnetic field, $B$, and the distance, $d$, between the excitation and ionisation regions are shown in Extended Data Fig.~\ref{fig:plasma}. The frequency of the observed oscillations as a function of the magnetic field is plotted in Extended Data Fig.~\ref{fig:plasma}b. The black solid line is a fit through the origin.
Previously, it has been shown that Rydberg EIT provides an effective probe of plasma formation \cite{Weller2016,Weller2019}.
Here we observe that, for any particular value of $d$, no oscillations are observed below a critical magnetic field, of order 4~G for the case of $d=6$~mm shown in Extended Data Fig.~\ref{fig:plasma}b. This suggests that a magnetic field is required in order to sustain a plasma \cite{Killian2007}, for long enough to induce significant ionisation. With no field, all charges are attracted to the wall, whereas with a magnetic field both the electrons and ions precess around the magnetic field vector. Above threshold, the frequency of the oscillations increase linearly with magnetic field \textit{B}. This suggests that the plasma decay rate, which determines the coefficient $\gamma$ in equation (\ref{eq:lv2}), increases linearly with $B$. This is consistent with higher precession frequencies, and hence higher collision rates at higher fields.

In addition in Extended Data Fig.~\ref{fig:plasma}c, we show the time dependence of the probe transmission for different value of the beam separations, $d$. This illustrates the switchover from transmission peaks for $d$ = 6~mm to dips at $d$ = 10~mm that we saw in the spectral data, Fig.~\ref{fig:spectra}. In Extended Data Fig.~\ref{fig:plasma}d we show how the linewidth of the pulses ($\Delta t_{1}$) and the time between them ($\Delta t_{2}$) vary with $d$. These measurements are performed at a fixed magnetic field, \textit{B} = 11.6~G. The duration of the pulse, $\Delta t_{1}$, increases with the separation $d$. We expect this as it takes a time---proportional to $d$---for the ground-Rydberg superpositions to travel to the ionisation region. Time delay effects have been a topic of considerable theoretical interest in Lotka-Volterra dynamics \cite{May1972,May1973}. Our experimental platform provides a useful system to provide a quantitative test of different models. We also observe that the period of the pulses decreased with $d$, although only by of order 20$\%$ over the distance range studied. This is also consistent with a faster plasma decay when the ionisation region is closer to the walls of the cell.

\subsection*{Lotka-Volterra magnetic field sensing}

\begin{figure}
\centering
\includegraphics[width=\linewidth]{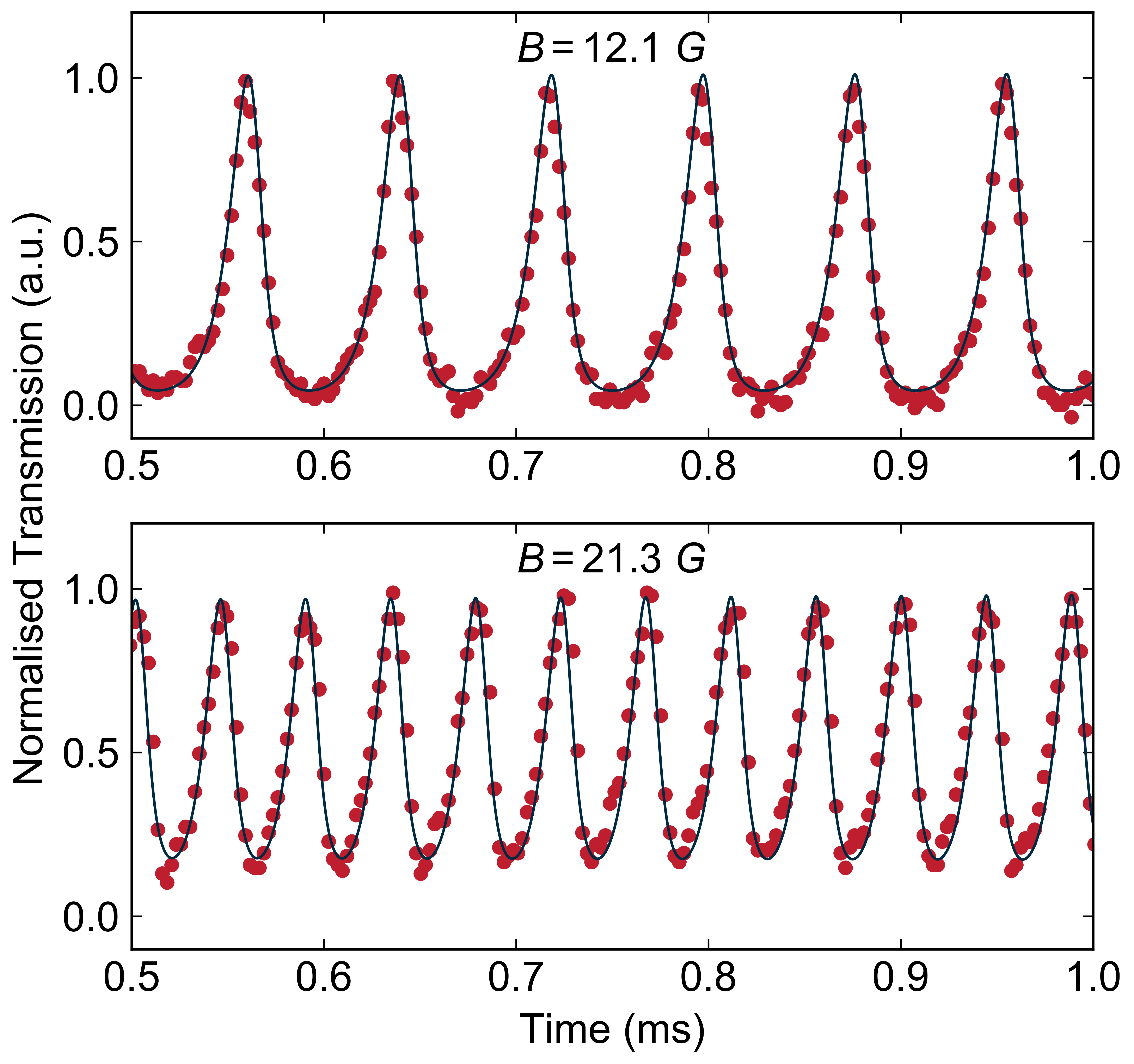}
\caption{\textbf{Lotka-Volterra magnetic field sensing.} Illustration of the dependence of the Lotka-Volterra cycle on the external magnetic field. For $B=12.1$~G we use Lotka-Volterra parameters $\alpha=0.651$, $\beta=0.217$, $\gamma=0.276$, and $\delta= 0.217$ in units of milliseconds. For $B=21.3$~G we use $\alpha=0.868$, $\beta=0.434$, $\gamma=0.566$, and $\delta= 0.434$ in units of milliseconds. In both cases we have used the condition $\beta=\delta$.}
\label{fig:LV_magnetic} \addtocounter{Sfigure}{1}
\end{figure}

In Extended Data Fig.~\ref{fig:LV_magnetic}, we illustrate how well the Lotka-Volterra model copes with changes in the real-world parameters, in this example, the external magnetic field, $B$. Note that we find that only changing the plasma decay constant, $\gamma$, is insufficient to reproduce the waveform. In further work, a thorough investigation of how each of the Lotka-Volterra parameters vary with each experimental parameter will be undertaken.

\textbf{Data availability} The data are available from the corresponding author on reasonable request.

\textbf{Acknowledgments}
This work was supported by the National Natural Science Foundation of China (No. 12241408, 12120101004, U2341211, 62175136 and 12222409), the National Key R\&D Program of China (grant No. 2020YFA0309400). C.S.A. acknowledges insightful discussions with Karen Wadenpfuhl. C.S.A. acknowledges financial support provided by the UKRI, EPSRC grant reference number EP/V030280/1 (“Quantum optics using Rydberg polaritons”).

\textbf{Author contributions} Y.J., Y.Z. and J.B. contributed to the building of the experimental setup, performed the measurements, these authors contributed equally to this work. C.S.A., J.Z., H.S. and Y.J. analysed the data. C.S.A. developed the theory. All authors discussed the results and contributed to the manuscript.

\textbf{Competing interests} The authors declare no competing interests.

\clearpage

\end{document}